\documentclass[prb,aps,preprint,showpacs]{revtex4}

\usepackage[english]{babel}
\usepackage[T1]{fontenc}
\usepackage[latin1]{inputenc}
\usepackage[dvips]{graphicx}
\usepackage{amsmath}
\usepackage{mathenv}
\usepackage{array}
\usepackage{bm}
\usepackage{fancyhdr}
\usepackage{makeidx}
\usepackage{subfigure}
\usepackage{babel}
\usepackage[v2]{xy}
\usepackage{epsfig}
\usepackage{amsfonts}
\usepackage{amssymb}

\begin{document}

\title{Quasi-Two-Dimensional Extraordinary Hall Effect} 
\author{N. Ryzhanova$^{1,2}$, A. Vedyayev$^{1,2}$, A. Pertsova$^2$, B. Dieny$^1$} 
\affiliation{
  $^1${\small SPINTEC, URA 2512 CEA/CNRS, 38054 Grenoble Cedex 9, France}\\
  $^2${\small Physics Department, M.V. Lomonosov Moscow State
University, Leninskie Gory, 1199991 Moscow, Russia}
  %$^*$corresponding author: vedy@magn.ru 
  }

\date{\today}

\begin{abstract}
Quasi-two-dimensional transport is investigated in a system consisting of 
one ferromagnetic layer placed between two insulating layers. Using 
the mechanism of skew-scattering to describe the Extraordinary Hall 
Effect (EHE) and calculating the conductivity tensor, we 
compare the quasi-two-dimensional Hall resistance with the resistance 
of a massive sample. In this study a new mechanism of 
EHE (geometric mechanism of EHE) due to non-ideal interfaces 
and volume defects is also proposed.
\end{abstract}
\pacs{75.47.-m, 75.70.-i, 73.50.-b}
\maketitle

\maketitle

\section{Introduction}

Recently there has been an increased focus on the 
fabrication of new spintronic devices based on the tunnel 
magnetoresistance effect  (TMR)\cite{Moodera1995} and the 
recently discovered spin torque effect (ST)\cite{Slonczewski1996} 
in multilayered structures consisting of several ferromagnetic 
nanolayers separated by thin insulating barriers. The geometric 
parameters of such structures also give us the possibility to 
investigate some of their quasi-two-dimensional transport 
properties and particularly to compare the relationship between 
diagonal and off-diagonal (responsible for EHE) conductivities 
in massive and above mentioned samples. This study is of great 
interest to the field of spintronics since quasi-two-dimensional 
EHE may provide an additional mechanism for recording and storing 
information in logic devices. Focusing on this aspect of EHE 
we study both diagonal and off-diagonal (Hall) conductivities 
in a system consisting of one ferromagnetic layer of 
thickness $a$ placed between two insulating layers and magnetized 
in the direction perpendicular to the interfaces ($z-$direction).

The remainder of the paper is organized as follows. In Section II 
EHE is attributed to the mechanism of skew-scattering\cite{Karplus1954} 
on the bulk and interface impurities. We report our results for 
the conductivity tensor including size effect terms calculated 
within the framework of the Kubo formalism and compare the 
quasi-two-dimensional Hall resistance $\rho^{H}_{2D}$ with 
that of a massive sample $\rho^{H}_{bulk}$. Section III   
discusses the nonideality of the interfaces  and the existence 
of volume defects that influence the form of current lines. 
We propose a new mechanism of EHE (we will refer to it 
as \textit{the geometric mechanism of EHE}) due to these 
defects using the diffusion equation\cite{Zhang2002} and 
taking into account that the diffusion coefficient has 
off-diagonal component proportional to the spin-orbit 
interaction. We summarize our results and offer some conclusions 
in Section IV.

\section{Skew-scattering mechanism of EHE in a three-layered structure \label{sec2}}

We will consider the geometry of current parallel to the 
$y-$direction resulting in the appearance of the $x-$component 
of the Hall field. In this case:
\begin{alignat}{2}\label{eq1}
j_{y} & =\sigma_{yy} E_{y} + \sigma_{yx} E^{H}_{x} & \qquad j_{x} & = \sigma_{xy} E_{y} + \sigma_{xx} E^{H}_{x}=0\nonumber\\
E^{H}_{x} & = -\frac{\sigma_{xy}}{\sigma_{xx}} E_{y} & \qquad \rho^{H}_{2D, bulk} & = \frac{\sigma_{yx}}{\sigma_{yy}\sigma_{xx} + \sigma_{yx}^{2}}
\end{alignat}
where $\sigma_{\alpha\beta}$ are diagonal and 
off-diagonal conductivities, $E^{H}_{x}$ is the Hall field. 
For the calculation of $\rho^{H}_{2D, bulk}$ we will use 
the Kubo formula with vertex corrections responsible for 
the transverse component of the current:
\begin{equation}\label{eq2}
\sigma_{\alpha \beta} = \frac{\hbar e^{2}}{\pi a^{4}_{0}}
\sum_{\kappa \kappa' z'} \nu_{\kappa \alpha}
\nu_{\kappa' \beta} G^{+}_{\kappa \kappa'}(z z') G^{-}_{\kappa' \kappa}(z'z)
\end{equation}
where $\vec{\nu_{\kappa}}$ is the velocity vector along the interface, 
$G^{+}_{0\kappa}(zz')$, $G^{-}_{0\kappa}(zz')$ are advanced and retarded 
Green's functions in mixed coordinate-momentum representation. To calculate 
the $y-x$ component of the conductivity tensor we will use the perturbation 
theory and will take into account only the corrections in linear order of 
the spin-orbit interaction:
\begin{align}
G^{+}_{\kappa \kappa'}(z z') = G^{+}_{0 \kappa}(z z') \delta_{\kappa \kappa'} 
+ G^{+}_{0 \kappa}(z z'') \sum_{\vec{\rho}}\left(T^{+}_{\kappa \kappa'}(\vec{\rho} z'') +
 H^{so}_{\kappa \kappa'}(\vec{\rho} z'')\right) G^{+}_{0 \kappa'}(z'' z')\label{eq3}\\
H^{so}_{\kappa \kappa'}(\vec{\rho} z'') = i \lambda^{so}\left(\vec{\rho} z''\right) 
a_{0}^{2} m_{z} \left[\kappa \times \kappa'\right]_{z} = i \lambda^{so} \left(\vec{\rho} z''\right)
a_{0}^{2} m_{z} \left(\kappa_{x} \kappa'_{y} - \kappa'_{x} \kappa_{y}\right)\label{eq4}
\end{align}
where $T^{+}_{\kappa \kappa'}(\vec{\rho} z)$ and $H^{so}_{\kappa \kappa'}(\vec{\rho} z)$ 
are the scattering matrix and the spin-orbit interaction,   correspondently, dependant 
on the type of atom in $(\vec{\rho} z)$ position ; $m_{z}$ is the unit vector along 
the magnetization, $\lambda^{so}_{m}$ is the spin-orbit parameter and $a_{0}$ is 
the lattice constant. It follows from Eq. (\ref{eq4}) that $H^{so}_{\kappa \kappa'}(\vec{\rho} z'') = -H^{so}_{\kappa'\kappa}(\vec{\rho} z'')$.

For the $T-$matrix in one-site approximation we can write down:
\begin{gather}
T_{\kappa \kappa'}(\vec{\rho} z) = \sum_{n} e^{i \left(\kappa - 
\kappa' \right) \left( \vec{\rho} - \vec{\rho_{n}}\right)} \frac{\epsilon_{n} - 
\Sigma \left(z\right)}{1 - \left(\epsilon_{n} - 
\Sigma \left(z\right)\right) G \left(\vec{\rho_{n}} z,\vec{\rho_{n}}z\right)}\label{eq5}\\
\lambda^{so}_{\kappa \kappa'} \left(\vec{\rho}z\right) = \sum_{m} e^{i \left(\kappa - 
\kappa'\right) \left(\vec{\rho} - \vec{\rho_{m}}\right)} \lambda^{so}_{m}
\label{eq6}
\end{gather}
where $\epsilon_{n}$ is the one-site energy. For the binary system $A B$ 
$\epsilon_{n}$ and $\lambda^{so}_{m}$ take values $\epsilon_{A,B}$ and $\lambda^{so}_{A,B}$, respectively; $G\left(\vec{\rho_{n}}z,\vec{\rho_{n}}z\right)=
\frac{a_{0}^{2}}{\pi} \int^{\frac{\sqrt{2 \pi}}{a_{0}}}_{0}\kappa d \kappa G_{\kappa} \left(z z\right)$. 
$\Sigma$ is the coherent potential and can be found from the system of self-consistent equations. 
The first equation of the system is valid for scattering on both bulk and interface impurities:
\begin{equation}\label{eq7}	
c_{A} \frac{\epsilon_{A} - \Sigma \left(z\right)}{1 - \left(\epsilon_{A} 
- \Sigma \left(z\right)\right) G_{0} \left(\vec{\rho_{n}} z,\vec{\rho_{n}}z\right)} + 
c_{B} \frac{\epsilon_{B} - \Sigma \left(z\right)}{1 - \left(\epsilon_{B} - \Sigma \left(z\right)\right) 
G_{0} \left(\vec{\rho_{n}} z,\vec{\rho_{n}} z\right)} = 0
\end{equation}
while the second one is written in the form corresponding to the interface scattring since we are interested in calculation of the interface coherent potential:
\begin{equation}\label{eq8}
G\left(\vec{\rho_{n}}z,\vec{\rho_{n}}z\right)=\frac{G_{0}\left(\vec{\rho_{n}}z,\vec{\rho_{n}}z\right)}{1-\Sigma\left(z\right)G_{0}\left(\vec{\rho_{n}}z,\vec{\rho_{n}}z\right)}
\end{equation}

For the calculation of $\sigma_{xx}$, $\sigma_{yy}$ we will use Eq. 
(\ref{eq2}) with Green's functions diagonal on $\kappa$ and renormalized 
on the coherent potential. For off-diagonal component averaging on the 
impurities distribution gives:
\begin{equation}\label{eq9}
	\sigma_{yx}=\frac{\hbar^{3}e^{2}M_{z}}{\pi a^{2}_{0} m^{2}}\sum_{\kappa
	\kappa'\vec{\rho_{n}}}\kappa^{2}_{y}\kappa'^{2}_{x} \left|G^{+}_{\kappa}
	\left(zz''\right)\right|^{2}\left|G^{+}_{\kappa}
	\left(z'z''\right)\right|^{2}Im\left(T^{+}_{\kappa\kappa'}(\vec{\rho_{n}}z'')
	\lambda^{so}_{\kappa'\kappa}\left(\vec{\rho_{n}}z''\right)\right)
\end{equation}
where $\vec{\rho_{n}}z''$ is the impurity position. We keep in Eq. 
(\ref{eq9}) only the main term with $n=m$.

For the binary $AB$ structure and purely random distribution of $A$, $B$ summing
 over $\vec{\rho_{n}}$ will give $\delta_{\kappa\kappa''}$. 
It is convenient to divide $\lambda^{so}$ into average and scattering parts:
\begin{gather}
\lambda^{so}_{A}=c_{A}\lambda^{so}_{A}+
c_{B}\lambda^{so}_{B}+
c_{B}\lambda^{so}_{A}-c_{B}\lambda^{so}_{B}=\bar{\lambda}+c_{B}\delta\lambda^{so}\label{eq10}\\
\lambda^{so}_{B}=c_{A}\lambda^{so}_{B}+c_{B}\lambda^{so}_{B}+
c_{A}\lambda^{so}_{A}-c_{A}\lambda^{so}_{A}=\bar{\lambda}-
c_{A}\delta\lambda^{so}
\label{eq11}
\end{gather}

The average in Eq. (\ref{eq9}) is:
\begin{equation}\label{eq12}
\begin{split}
&\frac{1}{N}\sum_{n}\left\langle T_{n}(z)\lambda^{so}_{n}(z)\right\rangle
\approx\\
&\quad \bar{\lambda}\left(c_{A}\frac{\epsilon_{A}-\Sigma\left(z\right)}
{1-\left(\epsilon_{A}-\Sigma\left(z\right)\right)G\left(\vec{\rho_{n}}z,
\vec{\rho_{n}}z\right)}
+
c_{B}\frac{\epsilon_{B}-\Sigma\left(z\right)}{1-
\left(\epsilon_{B}-\Sigma\left(z\right)\right)G\left(\vec{\rho_{n}}z,
\vec{\rho_{n}}z\right)}\right)\\
&+
\delta\lambda^{so}c_{A}c_{B}
\left(\frac{\epsilon_{A}-\Sigma\left(z
\right)}{1-\left(\epsilon_{A}-\Sigma\left(z\right)\right)G\left(\vec{
\rho_{n}}z,\vec{\rho_{n}}z\right)}
-
\frac{\epsilon_{B}-\Sigma\left(z
\right)}{1-\left(\epsilon_{B}-\Sigma\left(z\right)\right)G\left(\vec{
\rho_{n}}z,\vec{\rho_{n}}z\right)}\right)
\end{split}
\end{equation}

According to Eq. (\ref{eq7}) and Eq. (\ref{eq8}) 
we only take the imaginary part of the last term 
in Eq. (\ref{eq12}). We rewrite Eq. (\ref{eq12}) 
with renormalized Green's function:
\begin{equation}
\begin{split}
Im\frac{1}{N}\sum_{n}\left\langle T_{n}(z)\lambda^{so}_{n}(z)\right\rangle
\approx \delta\lambda^{so}c_{A}c_{B}Im&\left(
\frac{\left(\epsilon_{A}-
\Sigma\left(z\right)\right)\left(1-\Sigma\left(z\right)G_{0}
\left(\vec{\rho_{n}}z,\vec{\rho_{n}}z\right)\right)}{1-
\epsilon_{A}G_{0}\left(\vec{\rho_{n}}z,\vec{\rho_{n}}z\right)}\right.\\
&\left.-\frac{\left(\epsilon_{B}-\Sigma\left(z\right)\right)
\left(1-\Sigma\left(z\right)G_{0}\left(\vec{\rho_{n}}z,\vec{
\rho_{n}}z\right)\right)}{1-\epsilon_{B}G_{0}\left(\vec{
\rho_{n}}z,\vec{\rho_{n}}z\right)}\right)
\end{split}
\end{equation}\begin{equation}\label{eq14}
G_{0}\left(\vec{\rho_{n}}z,\vec{\rho_{n}}z\right)\equiv 
F_{0}\left(z\right)
\end{equation}

Eq. (\ref{eq5}) for coherent potential is:
\begin{equation}\label{eq15}
\Sigma\left(E,z\right)=\bar{\epsilon}+
\frac{c_{A}c_{B}\delta^{2}F\left(E,z\right)}{1-
\left(\tilde{\epsilon}-\Sigma\left(E,z\right)
\right)F\left(E,z\right)}
\end{equation}
\begin{align}\label{eq16}
\bar{\epsilon}=c_{A}\epsilon_{A}+c_{B}\epsilon_{B},\qquad 
\tilde{\epsilon}=c_{A}\epsilon_{B}+c_{B}\epsilon_{A},\qquad 
\delta=\epsilon_{A}-\epsilon_{B}
\end{align}

Usually it is convenient to choose $\bar{\epsilon}=0$. 
In this case $\tilde{\epsilon}=-\left(c_{A}-c_{B}\right)\delta$, $\epsilon_{A}=c_{B}\delta$, $\epsilon_{B}=-c_{A}\delta$.

Next we assume that scattering parameters for 
bulk impurities such as scattering potential 
and concentration are small enough to keep only 
the main terms for all values. In this case the 
imaginary part of $\Sigma$ is of order $\delta^{2}$ 
so for $\sigma^{bulk}_{yx}$ we will use:
\begin{equation}\label{eq17}
\begin{split}
\frac{1}{N}\sum_{n}\left\langle T_{n}(z)\lambda^{so}_
{n}(z)\right\rangle
&\approx
\delta\lambda^{so}_{bulk}\delta^{2}_{bulk}c_{A bulk}
c_{B bulk}\left(c_{A bulk}-c_{B bulk}\right)\\
&\times ImF\frac{1}{\left(1-c_{B bulk}\delta_{bulk}
ReF\right)^{2}\left(1+c_{A bulk}\delta_{bulk}ReF\right)^{2}}
\end{split}
\end{equation}

For the interface values the full self-consistent scheme 
is necessary. For both bulk and interface we suppose that 
the real part of the coherent potential just represents 
the renormalization of electron spectrum so $\Sigma$ can 
be considered as purely imaginary.

The zero order Green's function found from Schrödinger 
equation in $\kappa-z$ representation  is (we 
will further use the units with energy dimension $\left[L\right]=\dot{A}$):
\begin{equation}\label{eq18}
\begin{split}
G^{+}_{0\kappa}\left(0<z'<z<z_{1}\right)&=\frac{1}{2ik_{1}
\left(e^{ik_{1}a}\left(q+ik_{1}\right)^{2}-e^{-ik_{1}a}
\left(q-ik_{1}\right)^{2}\right)}\\
&\times\left(e^{ik_{1}\left(z-z_{1}\right)}\left(q-ik_{1}
\right)-e^{-ik_{1}\left(z-z_{1}\right)}\left(q+ik_{1}
\right)\right)\\
&\times\left(e^{ik_{1}z'}\left(q+ik_{1}
\right)-e^{-ik_{1}z'}\left(q-ik_{1}\right)\right)
\end{split}
\end{equation}      
\begin{align}
k_{1} & = \sqrt{\left(k^{\uparrow}_{F}\right)^{2}-\kappa^{2}+
i\frac{2k^{\uparrow}_{F}}{l_{1}}}\equiv c_{1}+id_{1}\label{eq19}\\
c_{1} & = \frac{1}{\sqrt{2}}\left(\sqrt{\left(\left(k^{\uparrow}_{F}
\right)^{2}-\kappa^{2}\right)^{2}+\frac{4k^{\uparrow}_{F}}{l_{1}}}+
\left(\left(k^{\uparrow}_{F}\right)^{2}-\kappa^{2}\right)\right)^{\frac{1}{2}}\label{eq20}\\
d_{1} & = \frac{1}{\sqrt{2}}\left(\sqrt{\left(\left(k^{\uparrow}_{F}\right)^{2}-
\kappa^{2}\right)^{2}+\frac{4k^{\uparrow}_{F}}{l_{1}}}-
\left(\left(k^{\uparrow}_{F}\right)^{2}-\kappa^{2}\right)\right)^{\frac{1}{2}}\label{eq21}\\
q & = \sqrt{q^{2}_{0}+\kappa^{2}},\qquad q^{2}_{0}=
\frac{2m}{\hbar^{2}}\left(U-E_{F}\right)\label{eq22}
\end{align}
where $0$ and $z_{1}$ are the coordinates of the left and 
right interfaces, $a$ is the layer thickness; 
$k^{\uparrow}_{F}$, $l_{1}$ are the Fermi momentum and 
the mean free path for spin "up", respectively, 
$ c_{1}d_{1}=\frac{\kappa^{\uparrow}_{F}}{l_{1}}$ 
(for spin "down" we will use index \textit{2}); $U$ 
is the height of the potential barrier.\par

The poles  of the Green's function in Eq. (\ref{eq18}) define 
the quantized energy spectrum of the thin ferromagnetic layer.

\subsection{Calculation of the bulk quasi-two-dimensional diagonal  conductivity}

For $\sigma_{xx}=\sigma_{yy}$ in Eq. (\ref{eq2}) we will take 
into account scattering on the interface responsible for size 
effect as well as on the bulk of the sample by using the Dyson 
equation with renormalized Green's function:
\begin{equation}\label{23}
G_{\kappa}\left(zz'\right)=G_{0\kappa}\left(zz'\right)+
G_{0\kappa}\left(z0\right)\Sigma G_{\kappa}\left(0z'\right)=
G_{0\kappa}\left(zz'\right)+\frac{G_{0\kappa}\left(z 0\right)
\Sigma G_{0\kappa}\left(0z'\right)}{1-G_{0\kappa}\left(00\right)\Sigma}
\end{equation}

Integrating over $z'$ from $0$ to $z$ for $z'<z$ and from $z$ 
to $a$ for $z'>z$ gives the conductivity in the units $ohm^{-1}cm^{-1}$:
\begin{equation}\label{eq24}
\sigma^{\uparrow}_{xx}=\frac{\sigma_{0} l_{1} 10^{8}}{2\pi k^{\uparrow}_{F}}
\int\frac{\kappa^{3} d\kappa Nom^{\uparrow}}{c_{1}Den^{\uparrow}}
\end{equation}
\begin{align}\label{eq24a}
\begin{split}
Nom^{\uparrow}&=\left(q^{2}+c^{2}_{1}\right)\left[\left(q^{2}+c^{2}_{1}+\left|\Sigma\right|^2+
2qRe\Sigma\right)\sinh{2d_{1}}a+2c_{1}\left|Im\Sigma\right|\cosh{2d_{1}}a\right]\\
&+2c_{1}Im\Sigma^{-}\left[\left(q^{2}+c^{2}_{1}\right)\cosh{2d_{1}}\left(z-a\right)+2\sinh^{2}d_{1}z\left(\cos 2c_{1}\left(z-a\right)+2qc_{1}\sin 2c_{1}\left(z-a\right)\right)\right]\\
&-2c_{1}\left(q+Re\Sigma\right)\left(q^{2}+c^{2}_{1}\right)\sinh 2d_{1}\left(z-a\right)\sin 2c_{1}z+\left(q^{2}+c^{2}_{1}+\left|\Sigma\right|^{2}+2qRe\Sigma\right)\\
&\times\left[\left(q^{2}+c^{2}_{1}\right)\sinh 2d_{1}\left(z-a\right)\cos 2c_{1}z-\sinh 2d_{1}z\left(\left(q^{2}-c^{2}_{1}\right)\cos 2c_{1}\left(z-a\right)+2qc_{1}\sin 2c_{1}\left(z-a\right)\right)\right]
\end{split}
\end{align}
\begin{align}\label{eq25}
\begin{split}
Den^{\uparrow}=\left(q^{2}+c^{2}_{1}\right)\left(\left(\left(q^{2}+c^{2}_{1}\right)+\left|\Sigma\right|^{2}+2qRe\Sigma\right)\cosh{2d_{1}a}+2c_{1}\left|Im\Sigma\right|\sinh{2d_{1}a}\right)\\
-\left[\left(q^{4}-6q^{2}c^{2}_{1}+c^{4}_{1}\right)+\left(\left|\Sigma\right|^{2}+2qRe\Sigma\right)\left(q^{2}-c^{2}_{1}\right)-4qc^{2}_{1}Re\Sigma\right]\cos2c_{1}a\\
-2qc_{1}\left(2\left(q^{2}-c^{2}_{1}+qRe\Sigma\right)+\left|\Sigma\right|^{2}\right)\sin{2c_{1}a}
\end{split}
\end{align}
where $\sigma_{0}=\frac{e^{2}}{2\pi\hbar}=\frac{10^{-3}}{13.6}(ohm^{-1})$ is the elementary conductivity of one channel.

For large enough layer thickness we can average Eq. (\ref{eq24}) over 
oscillations so that the averaged conductivity is:
\begin{equation}\label{eq26}
\left\langle \sigma^{\uparrow}_{xx}\right\rangle=\frac{\sigma_{0}l_{1}10^{8}}
{2\pi k^{\uparrow}_{F}}\int\frac{\kappa^{3}d\kappa}{c_{1}}\left[1-
\frac{l_{1}}{a}\frac{\left|Im\Sigma\right|c_{1}\sinh2d_{1}a}{\left(q^{2}+
c^{2}_{1}+\left|\Sigma\right|^{2}+2q Re\Sigma\right)\sinh2d_{1}a+2c_{1}
\left|Im\Sigma\right|\cosh2d_{1}a}\right]
\end{equation}

\begin{figure}
    \centering
        \includegraphics[width=10cm]{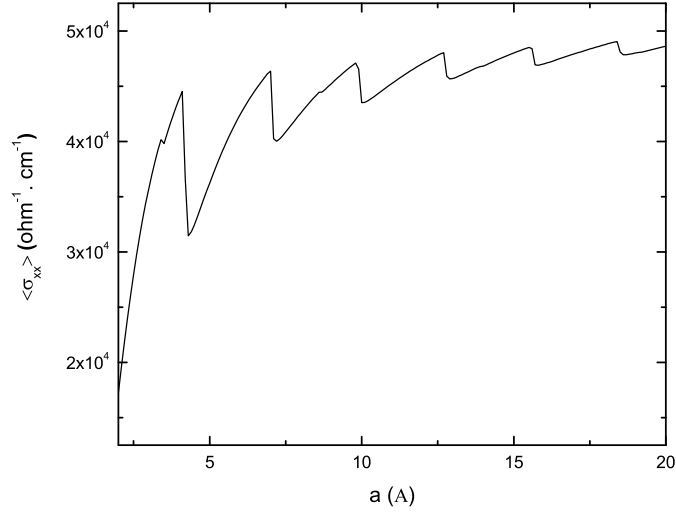}
        \caption{Averaged diagonal conductivity as a function
of $a$ (thickness): $<\sigma_{xx}>(a)$ 
for $k^{\uparrow}_{F}=1.1 (\dot{A}^{-1})$, $k^{\downarrow}_{F}=0.6 (\dot{A}^{-1})$, $l_{1}=100 (\dot{A})$, $l_{2}=60 (\dot{A})$, $c=0.3$ (see Eq. (\ref{eq26}))}\label{fig:asygmaxx}
\end{figure}

The first term in Eq. (\ref{eq26}) is the conductivity of the massive 
sample and the second one is due to the quasi-classical size effect. 
The full conductivity representing the sum of two spin channels is 
shown on Fig. (\ref{fig:asygmaxx}) as a function of the layer thickness.

\subsection{Calculation of the off-diagonal conductivity due to the spin-orbit interface scattering}

Now we will calculate $\sigma^{\uparrow}_{xy}(z)$ using 
Eq. (\ref{eq9}) with Green's function defined by Eq. (\ref{23}) 
and $z''=0$. After integration over $z'$ the conductivity is:
\begin{multline}\label{eq27}
\sigma^{\uparrow}_{xy}=
\frac{\sigma_{0}l_{1}a^{4}_{0}10^{8}}{2\pi^{2}k^{\uparrow}_{F}}\int\kappa^{3}d
\kappa c_{1}\frac{Im\left\langle T\left(\vec{\rho_{n}}0\right)\lambda^{so}\left(\vec{\rho_{n}}0\right)\right\rangle
\left(q^{2}+c^{2}_{1}\right)\sinh2d_{1}a}{\left|e^{ik_{1}a}\left(q+ic_{1}\right)^{2}\left(1+
\frac{\Sigma^{-}}{q+ic_{1}}\right)-e^{-ik_{1}a}\left(q-ic_{1}\right)^{2}\left(1+\frac{\Sigma^{-}}{q-ic_{1}}\right)\right|^{2}}\\
\times
\int\kappa^{3}d\kappa\frac{\left(q^{2}+c^{2}_{1}\right)\cosh2d_{1}\left(z-a\right)-\left(q^{2}-c^{2}_{1}\right)\cos2c_{1}\left(z-a\right)-2c_{1}q\sin2c_{1}\left(z-a\right)}{\left|e^{ik_{1}a}\left(q+ic_{1}\right)^{2}\left(1+\frac{\Sigma^{-}}{q+ic_{1}}\right)-e^{-ik_{1}a}\left(q-ic_{1}\right)^{2}\left(1+\frac{\Sigma^{-}}{q-ic_{1}}\right)\right|^{2}}
\end{multline}

This conductivity oscillates with the thickness and the distance 
from the interface $z=0$. Its behavior becomes more clear after 
averaging over oscillations:
\begin{equation}\label{eq28}
\begin{split}
\left\langle \sigma^{\uparrow}_{xy}\right\rangle & = 
\frac{\sigma_{0}l_{1}a^{4}_{0}10^{8}}{8\pi^{2}k^{\uparrow}_{F}}
\int\kappa^{3}d\kappa c_{1}\frac{Im\left\langle T\left(\vec{\rho_{n}}0\right)
\lambda^{so}\left(\vec{\rho_{n}}0\right)\right\rangle
\sinh2d_{1}a}{\left(q^{2}+c^{2}_{1}+\left|\Sigma^{-}\right|^{2}+
2qRe\Sigma^{-}\right)\sinh2d_{1}a+2c_{1}Im\Sigma^{-}\cosh2d_{1}a}\\
&\times
\int\kappa^{3}d\kappa\frac{\cosh2d_{1}\left(z-a\right)}{\left(q^{2}+c^{2}_{1}+
\left|\Sigma^{-}\right|^{2}+2qRe\Sigma^{-}\right)\sinh2d_{1}a+
2c_{1}Im\Sigma^{-}\cosh2d_{1}a}
\end{split}
\end{equation}

\begin{figure}
    \centering
        \includegraphics[width=10cm]{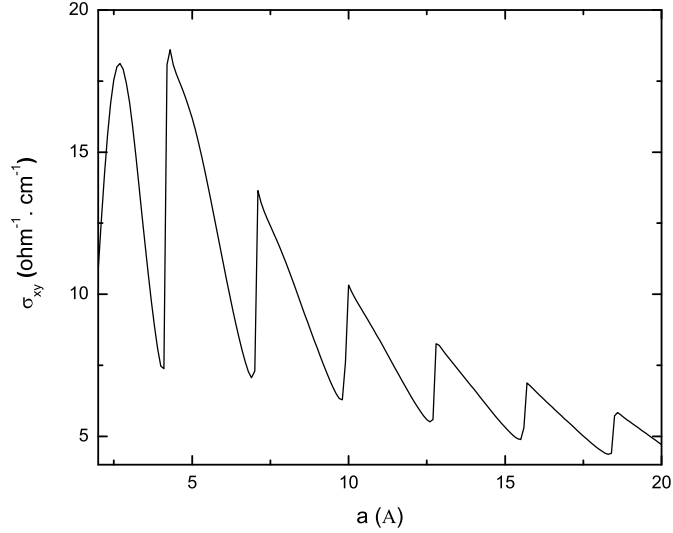}
\caption{Off-diagonal conductivity as a function
of $a$ (thickness): $\sigma_{xy}(a)$ for $k^{\uparrow}_{F}=1.1 (\dot{A}^{-1})$, 
$k^{\downarrow}_{F}=0.6 (\dot{A}^{-1})$, $l_{1}=100 (\dot{A})$, $l_{2}=60 (\dot{A})$, 
$c=0.3$, $\lambda^{so}=0.05 (\dot{A}^{-1})$ (see Eq. (\ref{eq27}))}\label{fig:sygmaxy}
\end{figure}

The same is done for spin "down". The sum of these two terms is shown 
on Fig. (\ref{fig:sygmaxy}).

It is clear that $\sigma^{\uparrow}_{xy}$ decreases with $z\rightarrow a$ since the functions 
$\cosh2d_{1}(z-a)$ have maximum values at $z=0$. Averaging $\sigma^{\uparrow}_{xy}$ over these functions gives the factor $\frac{l_{1}}{a}$ so for infinite $a$ this term tends to zero.

\begin{figure}
    \centering
        \includegraphics[width=10cm]{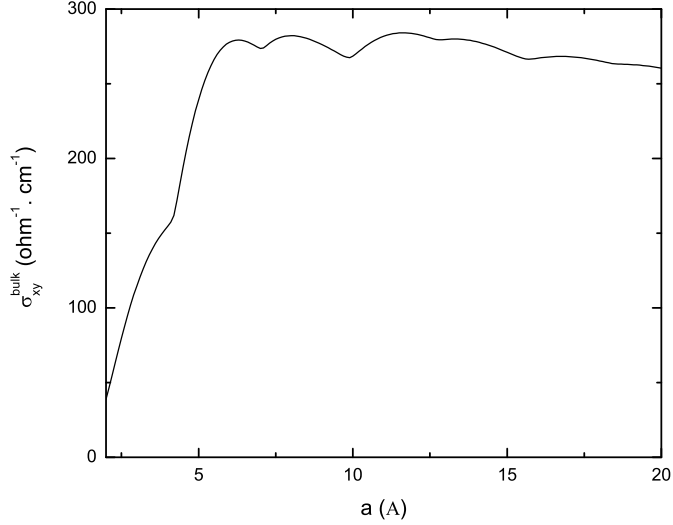}
\caption{Bulk off-diagonal conductivity as a function
of $a$ (thickness): $\sigma^{bulk}_{xy}(a)$ for $k^{\uparrow}_{F}=1.1 (\dot{A}^{-1})$, 
$k^{\downarrow}_{F}=0.6 (\dot{A}^{-1})$, $l_{1}=100 (\dot{A})$, $l_{2}=60 (\dot{A})$, $\lambda^{so}_{bulk}=0.03 (\dot{A})$, 
$c_{bulk}=0.01$, $\delta_{bulk}=1 (\dot{A}^{-1})$ (see Eq. (\ref{eq29}))}\label{fig:Bulk1}
\end{figure}

We also calculate the bulk conductivity $\sigma^{\uparrow}_{xy}+\sigma^{\downarrow}_{xy}$ 
(see Fig. (\ref{fig:Bulk1})) using Eq. (\ref{eq9}) with additional integration 
over $z''$ and bulk scattering parameters with  the bulk coherent potential in Bohrn approximation $\Sigma^{bulk}=ic_{bulk}(1-c_{bulk})\delta^{2}_{bulk}Im F^{bulk}(zz)$. 
In the absence of the interfacial scattering this approach gives us:
\begin{widetext}
\begin{align}\label{eq29}
\begin{split}
\sigma^{\uparrow bulk}_{xy}=\frac{\sigma_{0}l^{2}_{1}a^{3}_{0}Im\left
\langle T^{bulk}\lambda^{so}_{bulk}\right\rangle 10^{8}}{8\pi^{2}k^{\uparrow 2}_{F}}
\int\frac{\kappa^{3}d\kappa}{c_{1}}\frac{\left(q^{2}+c^{2}_{1}\right)^{2}\sinh2d_{1}a}{Den^{\uparrow}}
\times\int\frac{\kappa^{3}d\kappa}{c_{1}}\frac{1}{Den^{\uparrow}}\left\{\left(q^{2}+c^{2}_{1}\right)^{2}\right.\\
\left.\times
\sinh2d_{1}a-\left(q^{4}-c^{4}_{1}\right)\left(\sinh2d_{1}z\cos2c_{1}\left(z-a\right)-\sinh2d_{1}\left(z-a\right)\cos2c_{1}z\right)\right.\\
\left.
+2qc_{1}\left(\sinh2d_{1}z\sin2c_{1}\left(z-a\right)+\sinh2d_{1}\left(z-a\right)\sin2c_{1}z\right)\right\}
\end{split}
\end{align}
\end{widetext}
\begin{equation}
Den^{\uparrow}=\cosh2d_{1}a\left(q^{2}+c^{2}_{1}\right)^{2}-\left(q^{4}-6q^{2}c^{2}_{1}+c^{4}_{1}\right)\cos2c_{1}a+
4c_{1}q\left(q^{2}-c^{2}_{1}\right)\sin2c_{1}a
\end{equation}
or after averaging over oscillations:
\begin{equation}\label{eq30}
\sigma^{\uparrow bulk}_{xy}=\frac{\sigma_{0} l^{2}_{1} a^{3}_{0} Im\left\langle T^{bulk}
\lambda^{so}_{bulk}\right\rangle 10^{8}}{8\pi^{2}k^{\uparrow 2}_{F}}\left\{\int\frac{\kappa^{3} d
\kappa}{c_{1}}\right\}^{2}
\end{equation}

\begin{figure}
    \centering
        \includegraphics[width=10cm]{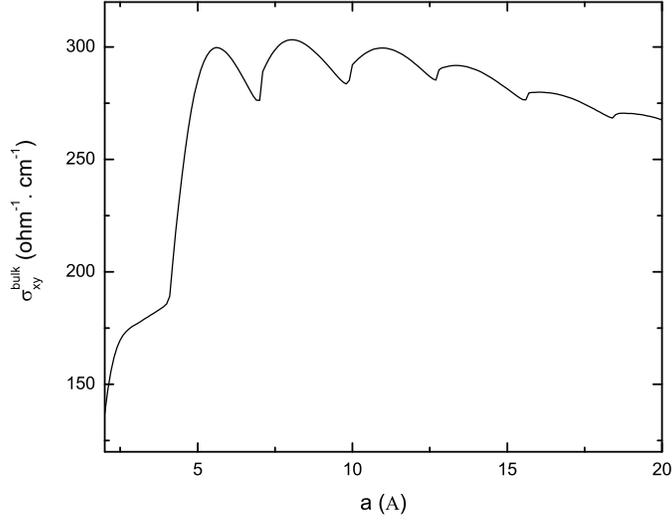}
\caption{Bulk off-diagonal conductivity with interfacial scattering as a function
of $a$ (thickness): $\sigma^{bulk}_{xy}(A)$ for $k^{\uparrow}_{F}=1.1 (\dot{A}^{-1})$, 
$k^{\downarrow}_{F}=0.6 (\dot{A}^{-1})$, $l_{1}=100 (\dot{A})$, $l_{2}=60 (\dot{A})$}\label{fig:Bulk2}
\end{figure}

\begin{figure}
    \centering
        \includegraphics[width=10cm]{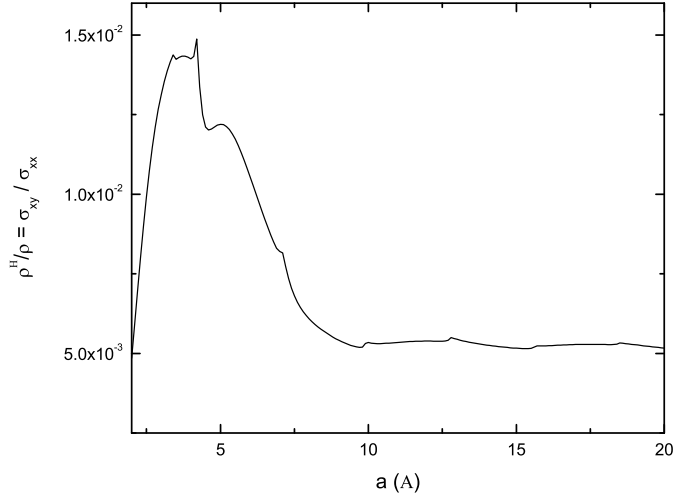}
\caption{Hall angle $\frac{\rho^{H}}{\rho}=\frac{\sigma^{xy}}{\sigma_{xx}}$ as a function
of $a$ (thickness); bulk parameters: $\lambda^{so}_{bulk}=0.03 (\dot{A}^{-1})$, $c^{bulk}=0.01$, 
$\delta^{bulk}=1 (\dot{A}^{-1})$; interface parameters: $\lambda^{so}=0.05 (\dot{A}^{-1})$, 
$c=0.3$, $\delta=1.5 (\dot{A}^{-1})$}\label{fig:Res}
\end{figure}

This conductivity has an oscillating behavior for the thin layer but tends to the constant 
value when $a\rightarrow\infty$ which coincides with its value for the massive sample. 
If we take into account the interfacial scattering the expression for $\sigma^{bulk}_{xy}$ becomes 
too complicated so we don't show it here. But the thickness dependences of $\sigma^{bulk}_{xy}$ 
and $\frac{\rho^{H}}{\rho}=\frac{\sigma_{xy}}{\sigma_{xx}}$ calculated using the full formula with 
renormalized Green's function are presented at Fig. (\ref{fig:Bulk2}) and Fig. (\ref{fig:Res}), 
correspondently. The bulk parameters are: $\lambda^{so}_{bulk}=0.03 (\dot{A}^{-1})$, $c^{bulk}=0.01$, 
$\delta^{bulk}=1(\dot{A}^{-1})$.

\section{Geometric mechanism of EHE \label{sec3}}

Let us consider an electric current through a thin ferromagnetic layer 
of thickness $a$ located between two thin insulating barriers and magnetized 
in $z-$direction perpendicular to the interfaces. The interfaces are not ideal 
and besides the impurities have topological defects which will be modeled as 
cylinders of radius $R$ so that the current lines in the vicinity of these defects 
follow their shape.

Diffusion equations for charge and spin currents in the absence of precession are:
\begin{align}
&\frac{\partial n}{\partial t}+\frac{\partial j^{x}_{e}}{\partial x}+
\frac{\partial j^{y}_{e}}{\partial y}= 0\label{eq31}\\
&\frac{\partial \vec{m}}{\partial t}+\frac{\partial j^{x}_{m}}{\partial x}+
\frac{\partial j^{y}_{m}}{\partial y}= -\frac{\vec{m}}{\tau_{sf}}\label{eq32}
\end{align}

For the stable state solution $\frac{\partial n}{\partial t}=\frac{\partial \vec{m}}
{\partial t}=0$. For currents we have the system of equations:
\begin{align}
j^{x}_{e} &= \sigma_{xx}E-D_{xx}\frac{\partial n}{\partial x}-
D_{xx}\beta'\frac{\partial m}{\partial x}-D_{xy}\frac{\partial n}{\partial y}-
D_{xy}\beta'\frac{\partial m}{\partial y}\label{eq33}\\
j^{y}_{e} &= \sigma_{yy}0+\sigma_{yx}E-D_{yy}\frac{\partial n}{\partial y}-
D_{yy}\beta'\frac{\partial m}{\partial y}-D_{yx}\frac{\partial n}{\partial x}-
D_{yx}\beta'\frac{\partial m}{\partial x}\label{eq34}\\
j^{x}_{m} &= \beta\sigma_{xx}E-D_{xx}\beta'\frac{\partial n}{\partial x}-
D_{xx}\frac{\partial m}{\partial x}-D_{xy}\beta'\frac{\partial n}{\partial y}-
D_{xy}\frac{\partial m}{\partial y}\label{eq35}\\
j^{y}_{m} &= \beta\sigma_{yy}0+\beta\sigma_{yx}E-D_{yy}\beta'\frac{\partial n}{\partial y}-
D_{yy}\frac{\partial m}{\partial y}-D_{yx}\beta'\frac{\partial n}{\partial x}-
D_{yx}\frac{\partial m}{\partial x}\label{eq36}
\end{align}

Here we take into account that $\vec{E}=\{E,0,0\}$, $\vec{m}=\{0,0,m\}$ are 
the electric field and spin accumulation correspondently, $D_{\alpha\beta}$ 
are the components of diffusion coefficient tensors. Off-diagonal components 
$D_{xy}$ and $D_{yx}$ of these tensors are proportional to the spin-orbit 
interaction and they are antisymmetrical in the $x-y$ transposition. For 
a metal with the cubic symmetry $\sigma_{xx}=\sigma_{yy}$, $D_{xx}=D_{yy}\equiv D_{0}$. 
Then we insert Eq. (\ref{eq33})-(\ref{eq36}) into Eq. (\ref{eq31}) and Eq. (\ref{eq32}) 
and after some manipulations we obtain two equations:
\begin{align}
&\Delta n= -\beta'\Delta m\label{eq37}\\
&\beta'\Delta n+\Delta m= \frac{m}{\tau_{sf}}\label{eq38}
\end{align}

And for $\tau_{sf}D_{0}(1-\beta'^{2})=\lambda^{2}_{sf}$:
\begin{equation}\label{eq39}
\Delta m-\frac{m}{\lambda^{2}_{sf}}=0
\end{equation}

For the cylindrical defect shape it is convenient to search for solution in 
polar coordinates so we can rewrite Eq. (\ref{eq39}):
\begin{equation}\label{eq40}
\frac{1}{r}\frac{\partial}{\partial r}\left(r\frac{\partial m}{\partial r}\right)+
\frac{1}{r^{2}}\frac{\partial^{2} m}{\partial \varphi^{2}}-
\frac{m}{\lambda^{2}_{sf}}\equiv \frac{\partial^{2} m}{\partial r^{2}}+
\frac{1}{r}\frac{\partial m}{\partial r}+
\frac{1}{r^{2}}\frac{\partial^{2} m}{\partial \varphi^{2}}-\frac{m}{\lambda^{2}_{sf}}=0
\end{equation}
where $\varphi$ is the angle between $x-$axe and the radius-vector $\vec{r}$ 
with the coordinates $(x, y)$.

The solution of Eq. (\ref{eq40}) is:
\begin{align}
&m=m_{1}\left(r\right)m_{2}\left(\varphi\right)\label{eq41}\\
&m_{2}\left(\varphi\right)=A_{1n}\cos n\varphi+A_{2n}\sin n\varphi\label{eq42}
\end{align}

As $\frac{\partial^{2} m_{2}}{\partial \varphi^{2}}=-m_{2}n^{2}$, 
Eq. (\ref{eq40}) can be transformed:
\begin{equation}\label{eq43}
m_{2}\left(\varphi\right)\left[\frac{\partial^{2} m_{1}}{\partial r^{2}}+
\frac{1}{r}\frac{\partial m_{1}}{\partial r}-\left(\frac{1}{\lambda^{2}_{sf}}+
\frac{n^{2}}{r^{2}}\right)m_{1}\right]=0
\end{equation} 

The solution of Eq. (\ref{eq43}) is\cite{Tikhonov1963}:
\begin{align}
&m_{1}\left(r\right)=B_{k} K_{k}\left(\frac{r}{\lambda_{sf}}\right)\label{eq44}\\
&m=\sum_{n}\left(A_{1n}\cos n\varphi+A_{2n}\sin n\varphi\right)K_{k}\left(\frac{r}{\lambda_{sf}}\right)
\label{eq44a}
\end{align} 
where $K_{k}(\frac{r}{\lambda_{sf}})$ is the solution of the modified 
Bessel equation\cite{Tikhonov1963}.

From Eq. (\ref{eq37}) it follows that:
\begin{align}\label{eq45}
&n=-\beta'm+n_{0}\\
&\Delta n_{0}=0
\end{align} 

For $n_{0}$ the solution is:     
\begin{equation}\label{eq46}
n_{0}=\sum_{n}\left(C_{1n}\cos n\varphi+C_{2n}\sin n\varphi\right)\frac{1}{r^{n}}
\end{equation} 

Taking into account Eq. (\ref{eq37}) we can rewrite Eq. (\ref{eq33}) and Eq. (\ref{eq34}):
\begin{align}\label{eq47}
j^{x}_{e} &= \sigma_{xx}E-D_{xx}\frac{\partial n_{0}}{\partial x}-
D_{xy}\frac{\partial n_{0}}{\partial y}\\
j^{y}_{e} &= \sigma_{yx}E-D_{yy}\frac{\partial n_{0}}{\partial y}+
D_{yx}\frac{\partial n_{0}}{\partial x}
\end{align} 

Now it is convenient to use the polar coordinate system and to write down $r-$ 
and $\varphi-$projections of the currents. Then we can use the boundary 
conditions to find unknown coefficients. These projection are:
\begin{equation}\label{eq48}
j^{0}_{re}=\sigma_{xx}E\cos\varphi+\sigma_{xy}E\sin\varphi
\end{equation}

(usual term)

\begin{equation}\label{eq49}
\delta j_{re}=-D_{xx}\frac{\partial n_{0}}{\partial x}\cos \varphi-
D_{xy}\frac{\partial n_{0}}{\partial y}\cos \varphi-
D_{yy}\frac{\partial n_{0}}{\partial y}\sin \varphi+
D_{xy}\frac{\partial n_{0}}{\partial x}\sin \varphi
\end{equation}

(additional diffusion term)

Now we will make some transformations:
\begin{align}\label{eq50}
&\frac{\partial n}{\partial x}=\frac{\partial n}{\partial r}
\frac{\partial r}{\partial x}+\frac{\partial n}{\partial \varphi}
\frac{\partial \varphi}{\partial x}\\
&\frac{\partial n}{\partial y}=\frac{\partial n}{\partial r}
\frac{\partial r}{\partial y}+\frac{\partial n}{\partial \varphi}
\frac{\partial \varphi}{\partial y}
\end{align} 

Using the expressions for derivatives of $r$, $\varphi$  over $x$, $y$, 
which is not too difficult to obtain, we write down the charge and spin 
currents in polar coordinates:  
\begin{equation}\label{eq51}
\begin{split}
&\delta j_{re}=-D_{0}\left[\cos\varphi\left(\frac{\partial n}{\partial r}
\frac{\partial r}{\partial x}+\frac{\partial n}{\partial \varphi}
\frac{\partial \varphi}{\partial x}\right)+\sin\varphi\left(\frac{\partial n}{\partial r}
\frac{\partial r}{\partial y}+\frac{\partial n}{\partial \varphi}
\frac{\partial \varphi}{\partial y}\right)\right]\\
&-D_{xy}\left[\cos\varphi\left(\frac{\partial n}{\partial r}
\frac{\partial r}{\partial x}+\frac{\partial n}{\partial \varphi}
\frac{\partial \varphi}{\partial x}\right)-\sin\varphi\left(\frac{\partial n}
{\partial r}\frac{\partial r}{\partial y}+\frac{\partial n}{\partial 
\varphi}\frac{\partial \varphi}{\partial y}\right)\right]\\
&=-D_{0}\frac{\partial n}{\partial r}-D_{xy}\frac{\partial n}{r\partial\varphi}
\end{split}
\end{equation}
\begin{align}
&\delta j^{n}_{rm}=-D_{0}\beta'\frac{\partial n}{\partial r}-
D_{xy}\beta'\frac{\partial n}{r\partial\varphi}\label{eq52}\\
&\delta j^{m}_{rm}=-D_{0}\frac{\partial m}{\partial r}-
D_{xy}\frac{\partial m}{r\partial\varphi}\label{eq53}\\
&\delta j^{0}_{rm}=\beta\sigma_{xx}E\cos\varphi+\beta\sigma_{yx}
E\sin\varphi\label{eq54}
\end{align} 

To find the unknown coefficients in Eq. (\ref{eq44a}) and Eq. (\ref{eq46}) 
we will use the boundary conditions on the surface of the cylinder 
representing that $r-$projection of currents are equal to zero:
\begin{equation}\label{eq55}
\begin{split}
&j^{n}_{R0}+\delta j^{n}_{R}=0\Rightarrow D_{0}\frac{\partial n_{0}}
{\partial r}|_{r=R}+D_{xy}\frac{\partial n_{0}}{R\partial\varphi}|_{r=R}=
-D_{0}\sum_{n}\left(C_{1n}\cos n\varphi+C_{2n}\sin n\varphi\right)\frac{n}{R^{n+1}}\\
&-D_{xy}\sum_{n}\left(C_{1n}\sin n\varphi-C_{2n}\cos n\varphi\right)\frac{n}{R^{n+1}}
\end{split}
\end{equation}

It gives us the system
\begin{align}\label{eq56}
&\sigma_{xx}E=-R^{-2}\left(D_{0}C_{11}-D_{xy}C_{21}\right)\nonumber\\
&\sigma_{xy}E=R^{-2}\left(D_{xy}C_{11}+D_{0}C_{21}\right)
\end{align}
with solution
\begin{align}\label{eq57}
&C_{11}=-\frac{D_{0}\sigma_{xx}-D_{xy}\sigma_{xy}}{D^{2}_{0}+D^{2}_{xy}}, 
C_{21}=\frac{D_{0}\sigma_{xy}+D_{xy}\sigma_{xx}}{D^{2}_{0}+D^{2}_{xy}} \nonumber\\
&n_{0}=-\frac{R^{2}E}{r}\frac{\left(D_{0}\sigma_{xx}-D_{xy}\sigma_{xy}\right)\cos\varphi-
\left(D_{0}\sigma_{xy}+D_{xy}\sigma_{xx}\right)\sin\varphi}{D^{2}_{0}+D^{2}_{xy}}
\end{align}

Spin current $r-$projection is also zero, and we can write down:
\begin{equation}\label{eq58}
\begin{split}
&j^{0m}_{R}+\delta j^{m}_{R}=0\Rightarrow \beta E\left(\sigma_{xx}\cos\varphi-
\sigma_{xy}\sin\varphi\right)=D_{0}\frac{\partial m}{\partial r}|_{r=R}+D_{xy}
\frac{\partial m}{R\partial\varphi}|_{r=R}\\
&=D_{0}\left(A_{11}\cos\varphi+A_{21}\sin\varphi\right)\frac{\partial}{\partial r}
K_{1}\left(\frac{r}{\lambda_{sf}}\right)|_{r=R}+D_{xy}\left(-A_{11}\sin\varphi+A_{21}
\cos\varphi\right)\frac{1}{R}K_{1}\left(\frac{R}{\lambda_{sf}}\right)
\end{split}
\end{equation}

Then we use some properties of Bessel functions\cite{Tikhonov1963}:
\begin{align}
&K_{1}\left(x\right)=\lim_{\nu\rightarrow 0}\frac{\pi}{2\sin\pi\left(\nu+1\right)}\left(I_{\nu-1}
-I_{\nu+1}\right)=-\lim_{\nu\rightarrow 0}\frac{\pi}{\sin\pi\nu}\frac{2\nu}{x}I_{\nu}\left(x\right)=
-\frac{I_{0}\left(x\right)}{x}\label{eq59}\\
&\frac{\partial K_{1}\left(x\right)}{\partial x}=\lim_{\nu\rightarrow 0}
\frac{\pi}{2\sin\pi\left(\nu+1\right)}\frac{\partial}{\partial x}\left(I_{\nu-1}-
I_{\nu+1}\right)=-\lim_{\nu\rightarrow 0}\frac{\pi}{\sin\pi\nu}\frac{2\nu}{x}
\frac{\partial}{\partial x}I_{\nu}\left(x\right)=-\frac{I_{1}\left(x\right)}{x}\label{eq60}
\end{align}
where $x\equiv\frac{r}{\lambda_{sf}}$, $\frac{\partial x}{\partial r}=
\frac{1}{\lambda_{sf}}$.  

Inserting Eq. (\ref{eq59})-(\ref{eq60}) in Eq. (\ref{eq58}) we get:
\begin{equation}\label{eq61}
\begin{split}
\beta E\left(\sigma_{xx}\cos\varphi-\sigma_{xy}\sin\varphi\right)&=-D_{0}\left(A_{11}\cos\varphi+
A_{21}\sin\varphi\right)\frac{I_{1}}{R}\\
&\quad+D_{xy}\left(A_{11}\sin\varphi-A_{21}\cos\varphi\right)
\frac{I_{0}\lambda_{sf}}{R^{2}}
\end{split}
\end{equation}
and it follows that:
\begin{align}\label{62}
&\beta E\sigma_{xx}=-D_{0}A_{11}\frac{I_{1}}{R}-D_{xy}A_{21}\frac{I_{0}\lambda_{sf}}{R^{2}}\\
&\beta E\sigma_{xy}=-D_{xy}A_{11}\frac{I_{0}\lambda_{sf}}{R^{2}}-D_{0}A_{21}\frac{I_{1}}{R}
\end{align}

\begin{align}\label{63}
&A_{11}=-\frac{\beta ER^{2}\left(\sigma_{xx}D_{0}I_{1}\left(\frac{r}{\lambda_{sf}}\right)R+
\sigma_{xy}D_{xy}I_{0}\left(\frac{r}{\lambda_{sf}}\right)\lambda_{sf}\right)}{\left(D_{0}
I_{1}\left(\frac{r}{\lambda_{sf}}\right)R\right)^{2}+
\left(D_{xy}I_{0}\left(\frac{r}{\lambda_{sf}}\right)\lambda_{sf}\right)^{2}}\\
&A_{21}=-\frac{\beta ER^{2}\left(\sigma_{xx}D_{xy}I_{0}\left(\frac{r}{\lambda_{sf}}\right)\lambda_{sf}-
\sigma_{xy}
D_{0}I_{1}\left(\frac{r}{\lambda_{sf}}\right)R\right)}{\left(D_{0}I_{1}\left(\frac{r}{\lambda_{sf}}\right)R\right)^{2}+
\left(D_{xy}I_{0}\left(\frac{r}{\lambda_{sf}}\right)\lambda_{sf}\right)^{2}}
\end{align}
\begin{equation}
\begin{split}
m=&-\frac{\beta ER^{2}\lambda_{sf}I_{0}\left(\frac{r}{\lambda_{sf}}\right)}{r\left\{\left(D_{0}
I_{1}\left(\frac{r}{\lambda_{sf}}\right)R\right)^{2}+
\left(D_{xy}I_{0}\left(\frac{r}{\lambda_{sf}}\right)\lambda_{sf}\right)^{2}\right\}}\\
&\times
\left\{\cos\varphi\left(\sigma_{xx}D_{0}I_{1}\left(\frac{r}{\lambda_{sf}}\right)R+\sigma_{xy}
D_{xy}I_{0}\left(\frac{r}{\lambda_{sf}}\right)\lambda_{sf}\right)\right.\\
&\quad\left.+\sin\varphi\left(\sigma_{xx}D_{xy}I_{0}\left(\frac{r}{\lambda_{sf}}\right)\lambda_{sf}-
\sigma_{xy}D_{0}I_{1}\left(\frac{r}{\lambda_{sf}}\right)R\right)\right\}
\end{split}
\end{equation}

Now we can define the additional Hall field due to  this cylindrical interface defect 
considering that Hall electrodes are the surfaces with 
coordinates $y=a$ and $y=-a$. This field is proportional to $n(a)-n(-a)$, 
$n=-\beta'm+n_{0}$, $r=\frac{a}{\left|\sin\varphi\right|}$. After integrating 
over $\varphi$ from $0$ to $\pi$ for the left surface and from $\pi$ to $2\pi$ 
for the right one we will have:
\begin{equation}\label{eq64}
n_{0}\left(a\right)-n_{0}\left(-a\right)=2\frac{ER^{2}}{a}\int^{\pi}_{0}d\varphi\frac{\sin^{2}
\varphi\left(D_{xy}\sigma_{xx}+D_{0}\sigma_{xy}\right)}{D^{2}_{0}+D^{2}_{xy}}=
\frac{\pi ER^{2}\left(D_{xy}\sigma_{xx}+D_{0}\sigma_{xy}\right)}{a\left(D^{2}_{0}+D^{2}_{xy}\right)}
\end{equation}

The second term due to $-\beta'm$ is:
\begin{equation}\label{eq65}
n_{m}\left(a\right)-n_{m}\left(-a\right)=\frac{2\beta\beta'ER^{2}}{a}\int^{\pi}_{0}d\varphi
\frac{\sin^{2}\varphi\left(\sigma_{xx}D_{xy}I_{0}\left(\frac{r}{\lambda_{sf}}\right)
\lambda_{sf}-\sigma_{xy}D_{0}I_{1}\left(\frac{r}{\lambda_{sf}}\right)R\right)}{\left(D_{0}I_{1}
\left(\frac{r}{\lambda_{sf}}\right)R\right)^{2}+\left(D_{xy}I_{0}\left(\frac{r}{\lambda_{sf}}\right)\lambda_{sf}\right)^{2}}
\end{equation}

At last, we have to multiply Eq. (\ref{eq64}) and Eq . (\ref{eq65}) by 
the concentrations of defects and electron charge. 

\section{Conclusion \label{sec4}}
It was shown that due to the additional scattering of electrons on the 
defects of the metal-insulator interfaces the total conductance decreases. 
From Eq. (\ref{eq26}) it follows that for small values of the ratio 
$\frac{a}{l}$ the bulk conductivity is completely suppressed and effective 
conductivity is proportional to the effective scattering length on the 
interfaces instead of the bulk mean free path. Hall conductivity, if we don't 
take into account the additional scew-scattering on the interface, 
decreases with decreasing the thickness of the ferromagnetic metallic layer. 
However the contribution to the Hall conductivity due to the additional 
scew-scattering on the interface increases. So the important characteristic 
of the considered device, Hall angle   $\frac{\sigma_{xy}}{\sigma_{xx}}=
\frac{\rho_{xy}}{\rho_{xx}}$, is larger for the thin ferromagnetic layer 
compared to the bulk layer. Besides that, the influence of insulator columns 
penetrating into the metallic layer may further increase the value 
of the Hall effect.

\begin{acknowledgments}
N. Ryzhanova and A. Vedyayev are grateful to SPINTEC for hospitality. This work 
was partially supported by the Russian Foundation for Basic Research.
\end{acknowledgments}

\end{document}